\documentclass[aps,pra,preprint,showpacs]{revtex4}
\usepackage{amssymb,amsmath,bm,graphicx}

\begin{document}
\bibliographystyle{apsrev}

\title{Corrections to the energy levels of a  spin-zero particle bound in a strong field}

\author{
R.N.~Lee $^{(a)}$\footnote{Electronic address: R.N.Lee@inp.nsk.su}, A.I.~Milstein$^{(a)}$\footnote{Electronic address: A.I.Milstein@inp.nsk.su}, S.G.~Karshenboim$^{(b,\,c)}$\footnote{Electronic address: sek@mpq.mpg.de}},

\affiliation{
$^{(a)}$Budker Institute of Nuclear Physics,
630090 Novosibirsk, Russia\\
$^{(b)}$D. I. Mendeleev Institute for
Metrology (VNIIM), St. Petersburg 198005, Russia\\
$^{(c)}$Max-Planck-Institut f\"ur Quantenoptik, 85748 Garching,
Germany\\
}
\date{\today}

\begin{abstract}
Formulas for  the corrections to the energy levels and wave functions of a spin-zero particle bound in a strong field are derived. General case of the sum of a Lorentz-scalar potential and  zero component of a Lorentz-vector potential is considered. The forms of the corrections differ essentially from those for spin-1/2 particles.
As an example of application of our results, we evaluated the
electric polarizability of a ground state of a spin-zero particle bound in a strong Coulomb field.
\end{abstract}

\maketitle

\section{Introduction}

As known, in many cases the perturbation theory (PT) is a very fruitful method to obtain analytic  results for various corrections. The formulas of PT for the Dirac equation  are similar to that for the Schr\"odinger equation and have  simple forms (see, e.g. \cite{BLP}). However, it is  essentially more complicated to derive the formulas of PT for the Klein-Gordon-Fock equation,  because this equation contains second derivative over time. In the present paper, we    derive the first and the second order corrections to the energy levels, and the first order correction to the wave function of a
spin-zero particle bound in a strong field. We consider a general case of a sum of the Lorentz-scalar and  the Lorentz-vector potentials. This may be useful at the consideration of the effects of the strong interaction in  pionic atoms. Using our formula for the correction to the energy level, we evaluate the
electric polarizability  of a ground state of a spin-zero particle bound in a strong Coulomb field.

\section{Perturbation theory}

The relativistic equation for the wave function of  a spin-zero particle (Klein-Gordon-Fock equation) bound  in the external time-independent field has  the form
\begin{equation}\label{eqKG}
\left[\left(i\frac{\partial}{\partial x^\mu}
-eA_{\mu}(\bf r)\right)^2-m^2-2mV(\bf r)\right]
\Psi({\bf r,\,t})=0\;,
\end{equation}
where $V(\bf r)$ is the Lorentz-scalar potential, $A_{\mu}(\bf r)$ is the Lorentz-vector potential, $e$ and $m$ are the charge and the mass of the particle, respectively; we set $\hbar=c=1$. The corresponding current, which obeys the continuity equation, reads
\begin{eqnarray}\label{J}
J_\mu= \Psi^*({\bf r ,\,t})\,i\frac{\stackrel{\leftrightarrow}{\partial}}
{\partial x^\mu}\,\Psi({\bf r,\,t})-2eA_{\mu}(\bf r)|\Psi({\bf r,\,t})|^2\,.
\end{eqnarray}
For ${\bm A}=0$, the solution with the fixed energy $E_n$,
 $\Psi({\bf r,\,t})=\exp(-iE_nt)\Phi_n({\bf r})$, obeys the equation
\begin{equation}\label{eqKGE}
\left[\left(E_n-U(\bf r)\right)^2-{\bf p}^2-m^2-
2mV(\bf r)\right]
\Phi_n({\bf r})=0\;,
\end{equation}
where $U({\bf r})=eA_0({\bf r})$. The non-relativistic approximation of this equation is the
Schr\"odinger equation with the  potential $U_{nr}({\bf r})=U({\bf r})+V({\bf r})$.   From
Eq.(\ref{J}) we find the normalization of the wave function
\begin{equation}\label{N}
2\int d{\bf r}\, [E_n-U({\bf r})]|\Phi_n({\bf r})|^2=1\, .
\end{equation}
Multiplying both sides of Eq.(\ref{eqKGE}) by $\Phi_k({\bf r})$
with $E_k\neq E_n$ and taking the integral over $\bf r$,  we obtain
\begin{equation}\label{normKG}
\int d{\bf r}\left[E_k+E_n-2U({\bf r})\right]
\Phi_k^*({\bf r})\Phi_n({\bf r})=0\;.
\end{equation}
Then we represent the potentials  in the form
$V({\bf r})=V_0({\bf r})+\delta V({\bf r})$, and
$U({\bf r})=U_0({\bf r})+\delta U({\bf r})$.
Let  $\phi_n({\bf r})$  be the solution   of the Klein-Gordon-Fock equation in the potentials $V_0({\bf r})$,  $U_0({\bf r})$ with the corresponding energy  $\varepsilon_n$.
Let us write
$$
\Phi_n({\bf r})=\phi_n({\bf r})+
\delta\phi_n({\bf r})\quad,\quad  E_n=\varepsilon_n+\delta\varepsilon_n\,.
$$
The first order correction to the wave function, $\delta\phi_n^{(1)}({\bf r})$, with respect to perturbations $\delta V({\bf r})$ and  $\delta U({\bf r})$ obeys
the equation
\begin{eqnarray}\label{delpsi}
&&\left[\left(\varepsilon_n-U_0({\bf r})\right)^2-{\bf p}^2-m^2-2mV_0({\bf r})\right]
\delta\phi_n^{(1)}({\bf r})\nonumber\\
&&=\left\{2m\delta V({\bf r})-2[\varepsilon_n-U_0({\bf r})]
[\delta\varepsilon_n^{(1)}-\delta U({\bf r})]\right\}
\phi_n({\bf r})\, .
\end{eqnarray}
Multiplying both sides of this equation by $\phi_n^*({\bf r})$,
 taking the integral over $\bf r$, and using normalization condition (\ref{normKG}),  we obtain
\begin{equation}\label{delE1}
\delta\varepsilon_n^{(1)}=\int d{\bf r}\, \{2m\delta V({\bf r})+
2\delta U({\bf r})[\varepsilon_n-U_0({\bf r)}]\}\,
|\phi_n({\bf r})|^2\, .
\end{equation}
Then we multiply Eq.(\ref{eqKGE}) by  $\phi_n^*({\bf r})$,
 take the integral over $\bf r$, and collect the terms of the second order in  $\delta V({\bf r})$ and
 $\delta U({\bf r})$. We have
 \begin{eqnarray}\label{delE2}
\delta\varepsilon_n^{(2)}&=&\int d{\bf r}\,
\{2m\delta V({\bf r})-2[\delta\varepsilon^{(1)}_n-
\delta U({\bf r})][\varepsilon_n-U_0({\bf r)}]\}\,
\phi_n^*({\bf r})\delta\phi_n^{(1)}({\bf r})\,\nonumber\\
&&-\int d{\bf r}\,
[\delta\varepsilon^{(1)}_n-
\delta U({\bf r})]^2\,|\phi_n({\bf r})|^2\, .
\end{eqnarray}
Let us introduce the Green's function  ${\cal D}_n({\bm r},\,{\bm r}')$, which obeys the
equation
\begin{eqnarray}\label{Green}
&&\left[\left(\varepsilon_n-U_0({\bf r})\right)^2-{\bf p}^2-m^2-2mV_0({\bf r})\right]{\cal D}_n({\bm r},\,{\bm r}')\nonumber\\
&&=\delta({\bm r}-{\bm r}')-\frac{1}{N}
\phi_n({\bf r})\phi_n^*({\bf r}')\, ,\nonumber\\
&&N_n=\int d{\bf r}\,|\phi_n({\bf r})|^2\, ,
\end{eqnarray}
and the conditions
\begin{equation}\label{conditions}
\int d{\bf r}'\,{\cal D}_n({\bm r},\,{\bm r}')
\phi_n({\bf r}')=0\, ,\quad
\int d{\bf r}\,\phi_n^*({\bf r}){\cal D}_n({\bm r},\,{\bm r}')
=0\, .
\end{equation}

It is obvious that ${\cal D}_n({\bm r},\,{\bm r}')$ is nothing but   the reduced Green function of the  Schr\"odinger equation, divided by $2m$,  with the effective energy
$\tilde E_n=(\varepsilon_n^2-m^2)/2m$, and the effective potential $\widetilde V({\bm r})=V_0({\bm r})+
(\varepsilon_n/m)U_0({\bm r})-U_0^2({\bm r})/2m$. We assume  that Eq.(\ref{Green}) can be solved. For instance, the solution is well known for the pure Coulomb field, $U_0({\bm r})=-Ze^2/r$ , $V_0({\bm r})=0$, where $Z$ is an atomic charge number and $e$ is the electron charge.

It is easy to check that the solution of Eq.(\ref{delpsi}) has the form
\begin{equation}\label{delpsisol}
\delta\phi_n^{(1)}({\bm r})=\int d{\bf r}'\,{\cal D}_n({\bm r},\,{\bm r}') \{2m\delta V({\bf r}')-2[\delta\varepsilon^{(1)}_n -
\delta U({\bf r}')][\varepsilon_n-U_0({\bf r}')]\}\,
\phi_n({\bf r}')+\beta\phi_n({\bf r}) \, ,
\end{equation}
where the constant $\beta$ is fixed by the normalization
condition (\ref{N}). We have
\begin{eqnarray}\label{beta}
\beta &=&-2\int\!\!\int\,d{\bf r} d{\bf r}'\, \phi_n^*({\bf r})[\varepsilon_n-U_0({\bf r)}]
{\cal D}_n({\bf r},{\bf r'})\nonumber\\
&&\times\{2m\delta V({\bf r}')-2[\delta\varepsilon^{(1)}_n-
\delta U({\bf r}')][\varepsilon_n-U_0({\bf r')}]\}\,
\phi_n({\bf r}')\,\nonumber\\
&&-\int d{\bf r}\,
[\delta\varepsilon^{(1)}_n-
\delta U({\bf r})]\,|\phi_n({\bf r})|^2\, .
\end{eqnarray}
Substituting
Eq.(\ref{delpsisol}) in Eq.(\ref{delE2}), we finally obtain

\begin{eqnarray}\label{delE2F}
\delta\varepsilon_n^{(2)}&=&\int\!\!\int\,d{\bf r} d{\bf r}'\,
\phi_n^*({\bf r})\{2m\delta V({\bf r})-2[\delta\varepsilon^{(1)}_n-
\delta U({\bf r})][\varepsilon_n-U_0({\bf r)}]\}
{\cal D}_n({\bf r},{\bf r'})\nonumber\\
&&\times\{2m\delta V({\bf r}')-2[\delta\varepsilon^{(1)}_n-
\delta U({\bf r}')][\varepsilon_n-U_0({\bf r')}]\}\,
\phi_n({\bf r}')\,\nonumber\\
&&-\int d{\bf r}\,
[\delta\varepsilon^{(1)}_n-
\delta U({\bf r})]^2\,|\phi_n({\bf r})|^2\, .
\end{eqnarray}
Note that the term $\beta\phi_n({\bf r})$ in Eq. ({\ref{delpsisol}) does not contribute to Eq.(\ref{delE2F})
due to Eq.(\ref{delE1}).

\section{Electric polarizability}

As an application of our results, let us consider the
electric polarizability $\alpha_{1s}$ of a ground state of a spin-zero particle
bound in a strong Coulomb field. For spin-$1/2$ particle the polarizability was obtained in Ref.\cite{ZMR,Yah}.

For the perturbation $\delta U({\bf r})=-e{\cal\bm E}\cdot {\bm r}$, we have $\delta\varepsilon^{(1)}=0$,  and $\delta\varepsilon^{(2)}$ can be  written in the form
\begin{equation}\label{alpha0}
 \delta\varepsilon^{(2)}=-\frac{1}{2}\alpha_{1s}{\cal E}^2
\, .
\end{equation}
From Eq.(\ref{delE2F}), we find
\begin{eqnarray}\label{alpha1}
\alpha_{1s}&=&\frac{2}{3}\alpha\,\left\{\int d{\bf r}\,
r^2\,\phi_{1s}^2(r)
-4\int\!\!\int\,d{\bf r} d{\bf r}'\,
\phi_{1s}( r)\left(\varepsilon_{1s}+\frac{Ze^2}{r}\right)
{\cal D}_{1s}({\bf r},{\bf r'})\right.\nonumber\\
&&\times\left. \left(\varepsilon_{1s}+\frac{Ze^2}{r'}\right)\,
\phi_{1s}( r') ({\bm r}\cdot{\bm r}')\right\}\, ,
\end{eqnarray}
where $\alpha=e^2=1/137$ is the fine structure constant.
The wave function and the energy of the ground state  has the form (see, e.g.,\cite{WF})
\begin{eqnarray}\label{phi1s}
\phi_{1s}( r)&=&\sqrt{A}\,(2\varkappa r)^{\gamma-1/2} \exp(-\varkappa r)\, ,\quad
\varepsilon_{1s}=m\sqrt{1/2+\gamma}\, ,\nonumber\\
\gamma&=&\sqrt{1/4-(Z\alpha)^2}\, ,\quad \varkappa=m\sqrt{1/2-\gamma}\, ,\quad
A=\frac{Z\alpha\varkappa^2}{\pi\Gamma(2\gamma+2)}\,,
\end{eqnarray}

Let us introduce the function ${\bf F}({\bf r})$
\begin{eqnarray}\label{F}
{\bf F}({\bf r})&=&\int\, d{\bf r}'\,
{\cal D}_{1s}({\bf r},{\bf r'})
\left(\varepsilon_{1s}+\frac{Ze^2}{r'}\right)\,
 {\bm r}'\,\phi_{1s}( r')\, .
\end{eqnarray}
It is convenient to represent it in the form
\begin{eqnarray}\label{F1}
{\bf F}({\bf r})&=&\frac{2\sqrt{A}\varepsilon_{1s}}{\varkappa^2}\,
(2\varkappa r)^{\gamma-3/2}\,\exp(-\varkappa r)\, g(2\varkappa r)\,{\bm r} \, ,
\end{eqnarray}
where the function $g(x)$ satisfies the equation, following from Eq.(\ref{Green}):

\begin{eqnarray}\label{eqg}
g''(x)+\left(\frac{2\gamma+1}{x}-1\right)\,g'(x)-\frac{2}{x^2}\,g(x)= \frac{x+1-2\gamma}{8}\,
.
\end{eqnarray}
Substituting Eqs.(\ref{F}) and (\ref{F1}) in Eq.(\ref{alpha1}), we obtain
\begin{eqnarray}\label{alpha2}
\alpha_{1s}&=&\frac{\alpha\varepsilon_{1s}}{3\varkappa^4}\Biggl\{
(1/2-\gamma)(\gamma+3/2)(\gamma+1)
\nonumber\\
&& - \frac{1}{\Gamma(2\gamma+1)} \int_0^\infty
dx\,(x+1-2\gamma)\,x^{2\gamma+1}\exp(-x)\,g(x) \Biggr\}\,.
\end{eqnarray}

The general solution  of Eq.(\ref{eqg}) has the form
\begin{eqnarray}\label{solg}
g(x)&=&-\frac{x}{4}-\frac{x^2}{16}+ \frac{x}{4}\,\,{}_2F_2\left[\left.\substack{
   \displaystyle 1,\, 1\\
   \displaystyle \gamma+2+\nu,\,\gamma+2-\nu}\right|x\right]\nonumber\\
&&+a\,x^{-\gamma+\nu}\,{}_1F_1\left[\left.\substack{
   \displaystyle -\gamma+\nu\\
   \displaystyle 1+2\nu}\right|x\right]
  +b\,x^{-\gamma-\nu}\,{}_1F_1\left[\left.\substack{
   \displaystyle -\gamma-\nu\\
   \displaystyle 1- 2\nu}\right|x\right]\,,
\end{eqnarray}
where ${}_pF_q$ is the hypergeometric function, $\nu=\sqrt{\gamma^2+2}$, and $a$ and $b$ are
some constants to be determined from the boundary conditions at $x=0$ and $x=\infty$. The condition at
$x=0$ gives $b=0$. The constant $a$ should be chosen to provide the     cancellation of  the exponentially large terms of
$g(x)$ at $x\to \infty$. The  large-$x$ asymptotics of the hypergeometric functions in Eq.
(\ref{solg}) can be calculated from their integral representations, see \cite{book}. The asymptotics have the form
\begin{eqnarray}
{}_2F_2\left[\left.\substack{
   \displaystyle 1,\, 1\\
   \displaystyle \gamma+2+\nu,\,\gamma+2-\nu}\right|x\right]&\backsim&
\Gamma[2+\gamma+\nu,2+\gamma-\nu]  \frac{\exp(x)}{x^{2\gamma+2}}
 \sum_n \frac{(1+\gamma+\nu)_n(1+\gamma-\nu)_n }{x^n n!}\nonumber\\
x^{-\gamma\pm \nu}{}_1F_1\left[\left.\substack{
   \displaystyle -\gamma\pm\nu\\
   \displaystyle 1\pm 2\nu}\right|x\right]&\backsim& \frac{\Gamma[1\pm
2\nu]}{\Gamma[\pm\nu-\gamma]}  \frac{\exp(x)}{x^{2\gamma+1}}
 \sum_n \frac{(1+\gamma+\nu)_n(1+\gamma-\nu)_n }{x^n n!}\,.
\end{eqnarray}
Thus, we obtain
\begin{eqnarray}\label{g}
g(x)&=&-\frac{x}{4}-\frac{x^2}{16}+ \frac{x}{4}\,\,
{}_2F_2\left[\left.\substack{
   \displaystyle 1,\, 1\\
   \displaystyle \gamma+2+\nu,\,\gamma+2-\nu}\right|x\right]
\nonumber\\
&&-\frac{\Gamma[2+\gamma+\nu,2+\gamma-\nu,\nu-\gamma]}{4\Gamma[1+
2\nu]}\,x^{-\gamma+\nu}\,{}_1F_1\left[\left.\substack{
   \displaystyle -\gamma+\nu\\
   \displaystyle 1+ 2\nu}\right|x\right]\,,
\end{eqnarray}

Using the integral representation of ${}_pF_q$, see \cite{book}, we obtain the following identity
\begin{eqnarray}\label{representation}
&&{}_2F_2\left[\left.\substack{
   \displaystyle 1,\, 1\\
   \displaystyle \gamma+2+\nu,\,\gamma+2-\nu}\right|x\right]
-\frac{\Gamma[2+\gamma+\nu,2+\gamma-\nu,\nu-\gamma]}{\Gamma[1+
2\nu]x^{1+\gamma-\nu}}{}_1F_1\left[\left.\substack{
   \displaystyle -\gamma+\nu\\
   \displaystyle 1+ 2\nu}\right|x\right]\nonumber\\
&=&\int_0^1 dz (1-z)^{\gamma+\nu}
   \int_0^1dt t^{\nu-\gamma-2}\exp[z x(1-1/t)]
\end{eqnarray}

Substituting Eqs. (\ref{representation}) and (\ref{g}) in Eq.(\ref{alpha2}) and integrating
over $x$, $z$, and $t$, we come to the final result for the polarizability $\alpha_{1s}$
\begin{eqnarray}
\alpha_{1s}
 &=&\frac{\alpha(\gamma +1) \sqrt{\gamma +1/2}}{3m^3(1-2 \gamma )}
   \left\{\frac{12 \gamma^2+64 \gamma +37}{2(1-2 \gamma) }\right.\nonumber\\
 &&-\frac{2 (2 \gamma +1)}{\gamma +\nu +1}
   \left(\frac{1-2 \gamma}{\gamma +\nu+2}\,
   {}_3F_2\left[\left.
   \substack{\displaystyle 1,2 \gamma +3,\gamma +\nu +1\\
   \displaystyle   \gamma +\nu +2,\gamma +\nu +3}\right|1\right]\right.\nonumber\\
 &&\left.\left.
   +\frac{(2 \gamma +3)}{\gamma +\nu   +3} \,
   {}_3F_2\left[\left.\substack{
   \displaystyle 1,2 \gamma +4,\gamma +\nu +1\\
   \displaystyle \gamma +\nu +2,\gamma +\nu +4}\right|1\right]\right)\right\}\,.
\end{eqnarray}
  The polarizability for arbitrary state can be obtained  similarly.
 Figure \ref{polar} shows the polarizability $\alpha_{1s}$
 in units $(m\alpha)^{-3}\,Z^{-4}$ as a function of $Z\alpha$.

\begin{figure}
  \setlength{\unitlength}{1cm}
  \includegraphics[width=7cm]{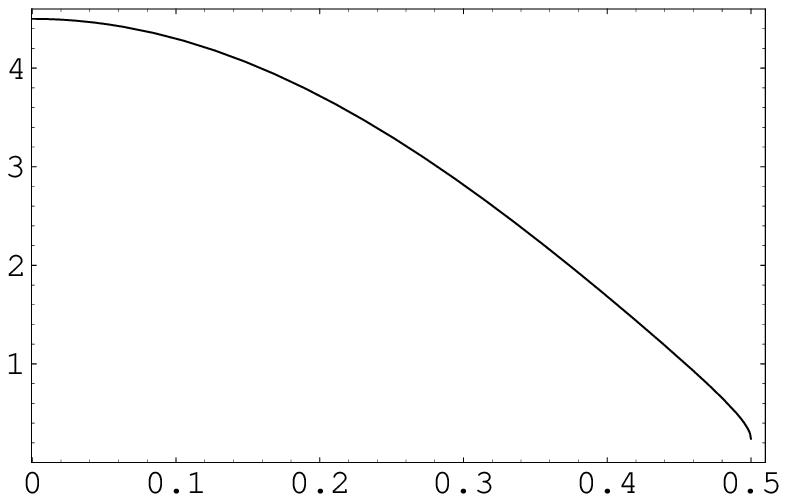}
\begin{picture}(0,0)
\put(-3.7,-0.5){$Z\alpha$} \put(-7.8,1.2){\rotatebox{90}{$(m\alpha)^{3}\,Z^{4}\alpha_{1s}$}}
\end{picture}

\caption{Polarizability in units $(m\alpha)^{-3}\,Z^{-4}$ as a function of $Z\alpha$ .}\label{polar}
\end{figure}

At  $Z\alpha\ll 1$,  we have
\begin{equation}\label{asymp}
\alpha_{1s}=\frac{9}{2(m\alpha)^3\,Z^4}\left[1-
\frac{121}{27}(Z\alpha)^2+\left(\frac{229}{72}+\frac{4 \pi ^2}{81}\right) (Z\alpha)^4+\ldots\right]\, .
\end{equation}
At $Z\alpha \rightarrow 1/2$ the polarizability has finite value $\sim 3.84\alpha/m^{3}$. Our
formula for polarizability can not be applied for $Z\alpha>1/2$ because we did not take into
account the effect of finite nuclear size. Besides, in a pionic atom it is necessary to account for the effects of the strong interaction that become dominant for the ground state at large
$Z$ .  It is interesting to compare the polarizabilities of a ground state,  
calculated for spin-$0$ particle and spin-$1/2$ particle. It turns out that the functions
$f(Z)=Z^4\alpha_{1s}$ obey the relation $f(Z)_{S=0}\approx f(2Z)_{S=1/2}$ with accuracy of a
few percent. 

In summary, we have derived the formulas for  the corrections to the energy levels and wave functions of a spin-zero particle bound in a strong field. These formulas may be useful at the consideration of various effects in pionic atoms. As an example,
we evaluated the
electric polarizability of a ground state of a spin-zero particle bound in a strong Coulomb field.

\section*{ACKNOWLEDGMENTS}
A.I.M. gratefully acknowledges the School of Physics at the University of New South Wales, and
the Max-Planck-Institute for Quantum Optics, Garching, for warm hospitality and financial
support during the visit. The work was supported in parts by RFBR Grants 03-02-16510,
03-02-04029, by DFG Grant GZ 436 Rus 113/769/0-1 and by Russian Science Support Foundation.

\end{document}